\documentclass[aps,pre,twocolumn,showpacs]{revtex4}
\usepackage{graphicx}
\usepackage{amsmath,amssymb}

\begin{document}

\title{Phase diagrams in the lattice RPM model: from order-disorder to gas-liquid phase 
transition}

\author{Alexandre Diehl\footnote{Corresponding author: diehl@fisica.ufc.br}} 
\affiliation{Departamento de F{\'\i}sica,
Universidade Federal do Cear{\'a}, Caixa Postal 6030, CEP
60455-760, Fortaleza, CE, Brazil}
\author{Athanassios Z. Panagiotopoulos} 
\affiliation{Department of Chemical Engineering, Princeton
University, Princeton NJ 08544}

\date{\today}
\begin{abstract}

The phase behavior of the lattice restricted primitive model (RPM)
for ionic systems with additional short-range nearest neighbor (nn) repulsive 
interactions has been studied by grand canonical Monte Carlo simulations. 
We obtain a rich phase behavior as the nn strength is varied. 
In particular, the phase diagram is very similar to 
the continuum RPM model for high nn strength. Specifically, we have found 
both gas-liquid phase separation, with associated Ising critical point, 
and first-order liquid-solid transition. We discuss how the line of 
continuous order-disorder transitions present
for the low nn strength changes into the continuum-space behavior
as one increases the nn strength and compare our findings with recent 
theoretical results by Ciach and Stell [Phys. Rev. Lett. {\bf 91}, 060601 (2003)].
\end{abstract}


\maketitle

\section{Introduction}

In spite of the considerable progress made in the last decade in getting 
a clear picture of critical behavior and phase separation in systems 
dominated by Coulombic interactions, ionic systems remain the subject of 
intense interest. For a symmetric 1:1 electrolyte system, for example, recent 
experiments~\cite{Wiegand97,Wiegand981,Wiegand982,Weingartner01,Bianchi01} 
and simulations~\cite{pablo01,azp02,caillol02,Luijten02,Kim04} strongly support
three-dimensional Ising-like criticality as the asymptotic
behavior. One of the most basic and successful models of ionic fluids is the 
restricted primitive model (RPM), in which the ions are viewed as 
equisized hard spheres carrying positive and
negative charges of the same magnitude. Used for both theoretical and 
Monte Carlo simulations, the RPM model is able to characterize properly 
the vapor-liquid phase transition observed in electrolyte 
solutions~\cite{Fisher94,Stell95,Fisher93}, as well the solid-liquid transitions 
of molten salts~\cite{Vega03}.

In recent years, in order to understand better criticality in the RPM
model, a lattice version of this model has been
introduced~\cite{Dickman99,azp99,Ciach,Ciach03,Ciach04,Kobelev02,Kobelev03,Diehl03,Fisher04}. 
In this model, the positions of the positive and negative ions are
restricted to the sites of an underlying lattice with a spacing equal to the ionic diameter. 
The most striking feature of this model is the presence of an order-disorder 
transition, which is absent in the continuous version of the RPM. There is no 
gas-liquid transition and the coexistence is between a low-density disordered 
phase and an antiferromagnetically ordered high-density phase; the 
transition is continuous (N\'eel-type line) above and first-order below 
a tricritical point. By contrast, non-ionic fluids have the same critical 
behavior as the lattice Ising model such that continuum and lattice models are 
essentially equivalent~\cite{azp00}. 

Although the presence of an underlying lattice naturally favors
the appearance of charge ordering, it is not completely obvious why the
lattice and continuum models of RPM present such a different
critical behavior. A possible explanation has been advanced by
Ciach and Stell~\cite{Ciach,Ciach03,Ciach04} using a formalism based on the
Landau-Ginzburg-Wilson approach. They proposed that in contrast to
the uncharged systems with short-range interactions, where the
long-wavelength fluctuations dominate and the lattice structure is
irrelevant, in ionic systems the short-wavelength charge
fluctuations are the most important. In this case, the
short-distance properties of the system, such as the lattice
structure or the shape of the short-range potentials added to the
RPM model, become important and different phase diagrams can be
obtained, i.e. there is no universality. In fact, Ciach and Stell
have predicted that for a model system with additional short-range
interactions added to the RPM model both gas-liquid and
tricritical points can be thermodynamically stable. 
Also, more recently~\cite{Ciach03,Ciach04} they proposed that when repulsive 
nearest neighbor interactions are included to the lattice RPM model, 
the phase diagram obtained should be qualitatively the same as in the 
continuum RPM model. These short-range interactions could represent the 
interaction between the ions and the particles of the solvent in 
which the ions are dissolved.

In this paper we extend the lattice RPM model for ionic systems introduced 
in Ref.~\cite{Diehl03}, where a short-range attractive potential was 
supplemented to the lattice RPM. Now, in addition to the hard-core and 
electrostatic interactions, some short-range repulsive interactions 
between the ions are included. We use grand canonical Monte Carlo 
simulations, combined with histogram reweighting~\cite{Ferrenberg88} 
and mixed-field finite-size scaling~\cite{Wilding92} 
techniques, to obtain the coexistence curves and the associated 
critical points. The paper is organized as follows. The model and 
the computational details are given in Sec.~\ref{model}. 
Results are discussed in Sec.~\ref{results}. We close in 
Sec.~\ref{conclusions} with summary and conclusions.

\begin{figure}
\includegraphics[width=3.5cm]{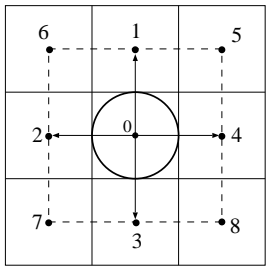}
\caption{Two-dimensional projection of the three-dimensional lattice 
structure used in our simulations. 
In addition to the electrostatic potential, a charged particle in site 0 will 
be affected by a repulsion with the other particles on the first nearest neighbor 
sites (1, 2, 3 and 4, plus two more off the plane). For the second 
nearest neighbor sites (5, 6, 7 and 8, and the corresponding off-plane positions) 
just the electrostatic potential is considered.} 
\label{fig1}
\end{figure}

\section{Model system and simulation methods}
\label{model}

The model used here is essentially the same as that of Ref.~\cite{Diehl03}, 
but with repulsive rather than attractive interactions. 
We consider a system of $2N$ charged hard spheres of equal diameter $\sigma$, 
half of them carrying charge $+q$ and half charge $-q$, interacting through 
the pair potential
\begin{equation}
\label{potential}
U_{ij}= \begin{cases}
\frac{\displaystyle q_i q_j}{\displaystyle D r_{ij}} & \text{if $r_{ij} \ge \sigma\;,$}\\
+\infty & \text{if $r_{ij} < \sigma\;,$}
\end{cases}
\end{equation}
where $D$ is the dielectric constant of the structureless solvent
in which the ions are immersed. In addition to the electrostatic potential and 
hard-core given in Eq.~(\ref{potential}), we include a short-range 
repulsive potential of strength $J > 0$ between the first (nn) nearest neighbor ions, 
regardless of their charge. For $J=0$ the lattice RPM model is recovered. 
These ions are restricted to the sites of a 
three-dimensional simple cubic lattice with a unit cell length $l$. 
In Fig.~\ref{fig1} we show a two-dimensional projection of the lattice 
structure used in our simulations. 
Reduced quantities are defined as follow
\begin{equation}
\label{reduced} T^{\ast}= \frac{\kappa_B T}{E_0}, \quad
J^{\ast} = \frac{J}{E_0}, \quad \mbox{and}
\quad \rho^{\ast}=\frac{2N\sigma^3}{V}\;,
\end{equation}
where $\sigma$ is the ion diameter, $V$ is the volume of the system and 
$E_0=q^2/D\sigma$ is the Coulomb energy between two ions at close contact. 
The reduced chemical potential, $\mu^{\ast}$, is defined so that at the 
limit of high temperatures and low densities, 
$\mu^{\ast}\to 2T^{\ast}\ln N\sigma^3/V$, where the factor 2 comes from the 
presence of two ions per minimal neutral ``molecule'' inserted or 
deleted in the simulations. Using these definitions, the effect of 
the repulsive interactions on the 
properties of the lattice RPM model can be monitored by changing the reduced 
energy parameter, $J^{\ast}$.

The simulations were performed using the discretization methodology 
introduced by Panagiotopoulos and Kumar~\cite{azp99}.
In this approach the allowed positions for the centers of the ions
are on a simple cubic grid of characteristic length $l$. Also,
the lattice discretization parameter is defined as $\zeta =
\sigma/l$, such that the lattice and continuum limits can be
reproduced by changing $\zeta$. In fact for $\zeta =10$ the 
results were nearly indistinguishable from the continuum model and 
the critical parameters of the RPM were well reproduced~\cite{azp02,Romero00}.
In this work we study the lattice RPM model corresponding to $\zeta =1$. 
The Ewald sums were performed with conducting boundary conditions, 
using 518 Fourier-space wave vectors and real-space damping 
parameter $\kappa =5$.

We used grand canonical Monte Carlo (GCMC) simulations with pair
additions and removals at each time step. To enhance acceptance 
of the insertion and removal steps we used
distance-biased sampling, introduced in Ref.~\cite{orkoulas94}. 
Multihistogram reweighting~\cite{Ferrenberg88,Frenkel,azp002} 
techniques were used to analyze the simulation data. 
For the critical region we used mixed-field finite size scaling
(FSS) analysis proposed by Bruce and Wilding~\cite{Wilding92},
which accounts for the lack of symmetry between coexisting phases
in fluids. We did not attempt to incorporate corrections for 
pressure mixing in the scaling fields, as any such effects are 
expected to be small~\cite{Young04}. 
Typical runs involve $2-5\times 10^7$ Monte Carlo steps (MCs) for
equilibration and $2-9\times 10^8$ MCs for production.
Statistical uncertainties for the critical parameters were
produced from 8 to 24 independent runs, depending on system size,
at near critical conditions, with different seeds used for 
the random number generator routine ``ran2'' of Ref.~\cite{recipes}.

\section{Results and discussion}
\label{results}

In this section we present our results obtained for the lattice RPM 
model with nn repulsion. Since $J^{\ast}=0$ represents the pure lattice 
RPM model, for which the phase diagram is known, we begin our discussion 
analyzing the dependence of the density on the chemical potential 
when we increase the nn repulsive strength $J^{\ast}$. 

\begin{figure}
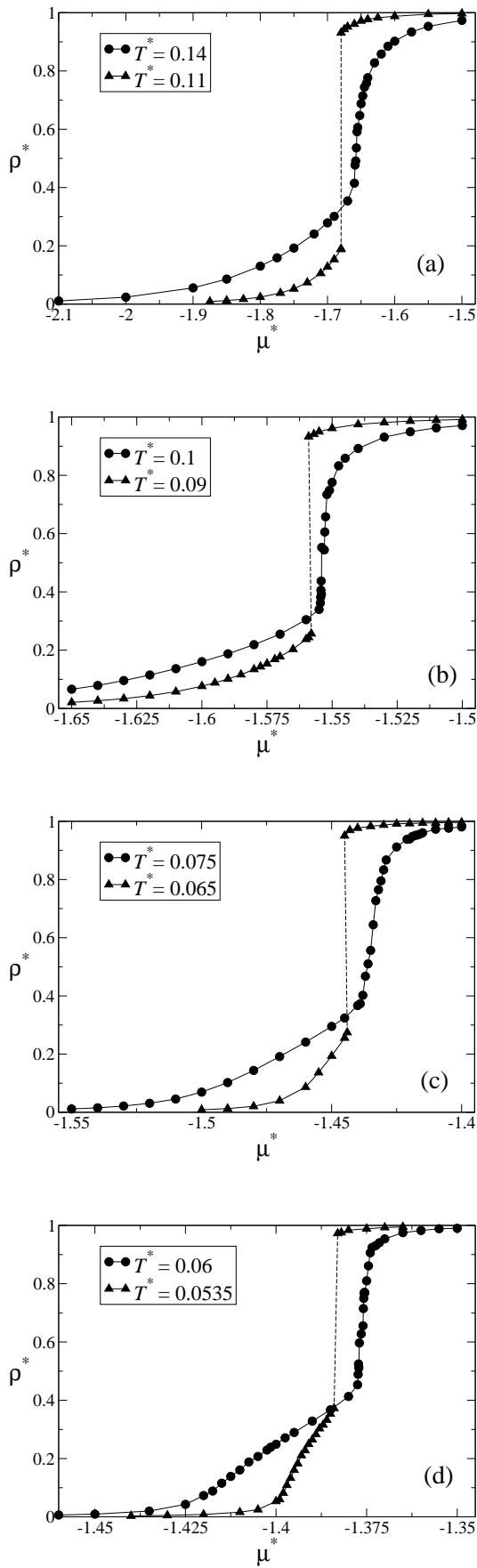

\includegraphics[width=7cm]{fig2a.eps}{\vspace*{.8cm}}
\includegraphics[width=7cm]{fig2b.eps}{\vspace*{.8cm}}
\includegraphics[width=7cm]{fig2c.eps}{\vspace*{.8cm}}
\includegraphics[width=7cm]{fig2d.eps}
\caption{Isotherms calculated in grand canonical simulations ($L^{\ast}=12$) 
for different repulsive strength, (a) $J^{\ast}=0.01$, (b) 0.03, 
(c) 0.05 and (d) 0.06. The dashed line marks the approximated 
location of the order-disorder phase transition. The solid lines are just 
guides to the eye. Statistical uncertainties are smaller than the symbol size.} 
\label{fig2}
\end{figure}

\begin{figure}
\includegraphics[width=3.5cm]{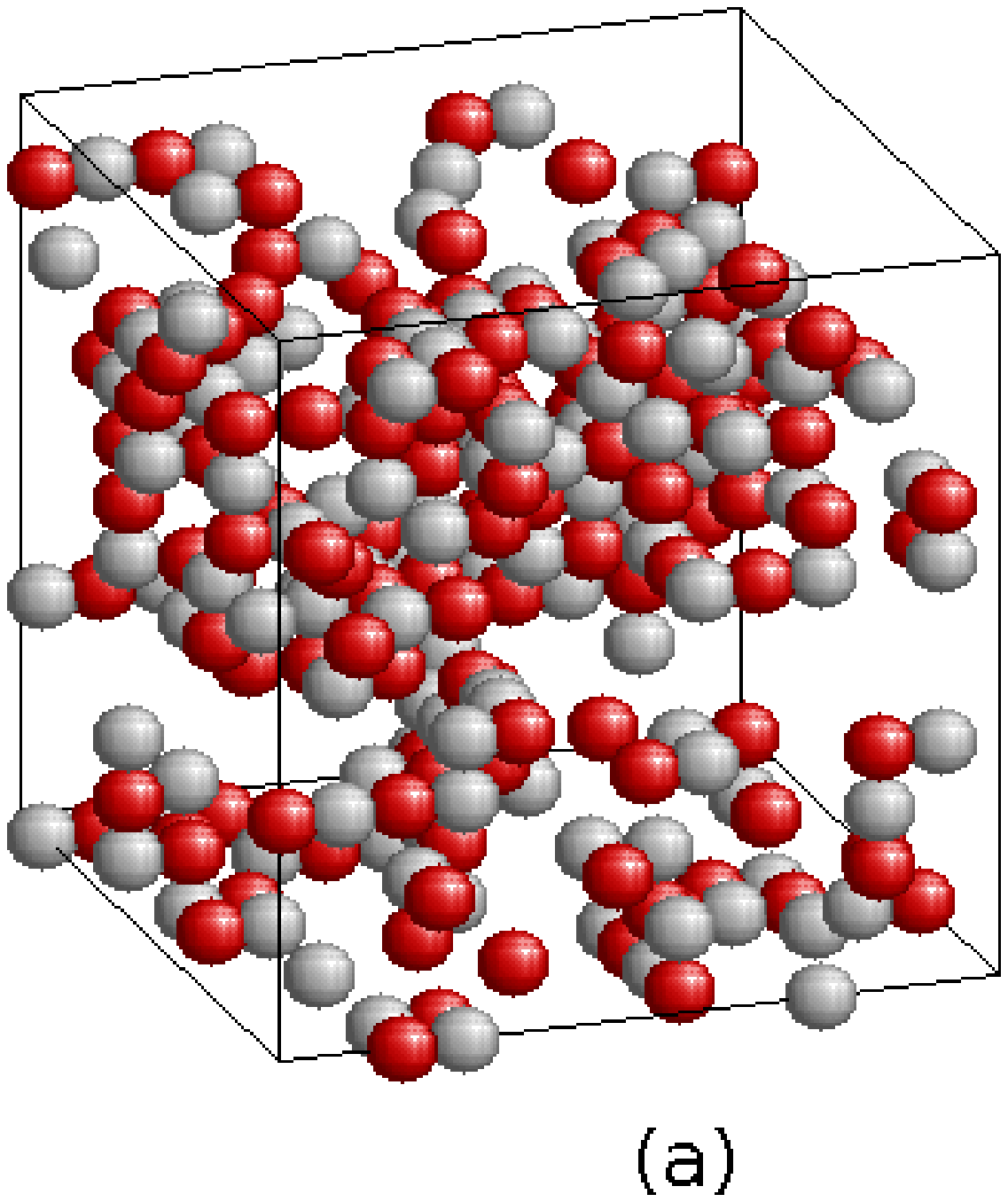}{\hspace*{.8cm}}
\includegraphics[width=3.5cm]{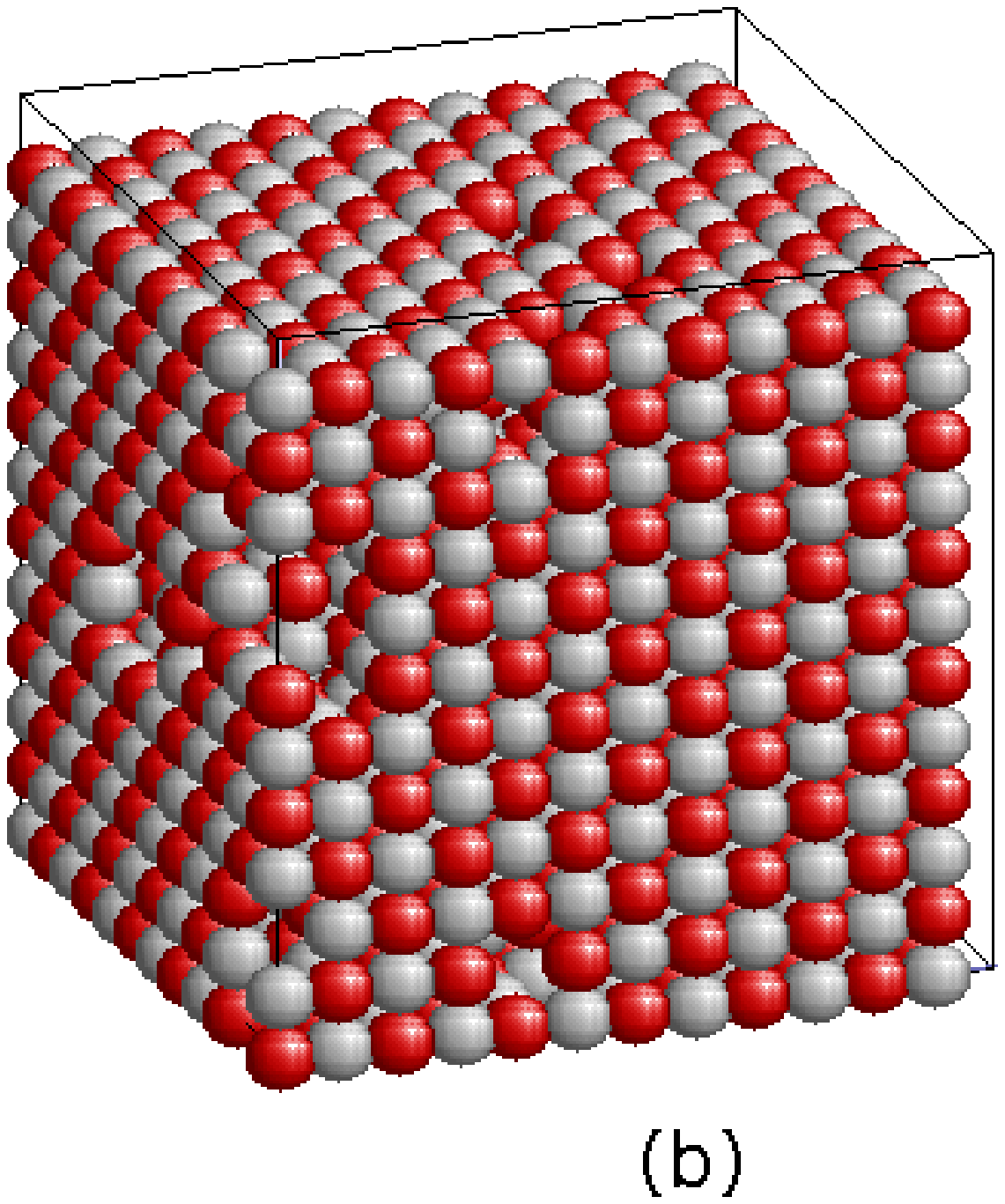}
\caption{\label{fig3} (color online) Coexisting phases generated for 
$J^{\ast}=0.01$. The temperature is $T^{\ast}=0.11$ and the densities 
are (a) $\rho^{\ast}=0.19$ and (b) $\rho^{\ast}=0.93$.} 
\end{figure}

\begin{figure}[h]
\vspace*{.7cm}
\includegraphics[width=8cm]{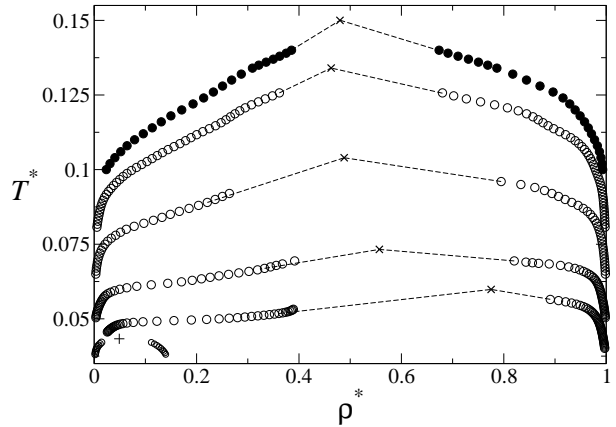}
\caption{\label{fig4} Phase diagram as a function of the repulsive strength $J^{\ast}$. 
Points (open circles) from top to bottom are for 
$J^{\ast}$= 0.01, 0.03, 0.05, 0.06, and 0.3, respectively. 
The $\zeta =1$ lattice RPM results of Panagiotopoulos and 
Kumar~\cite{azp99} ($J^{\ast}=0$), are shown as filled circles. 
Tricritical points ($\times$) were obtained using a linear extrapolation (dotted 
lines) of coexistence lines, as expected for $d=3$ tricriticality. The 
critical point ($+$) for $J^{\ast}=0.3$ was estimated from finite size scaling 
using $L^{\ast}=15$. Statistical uncertainties are smaller than the symbol size.}
\end{figure}

Figure \ref{fig2} shows some isotherms calculated for different 
repulsive strength $J^{\ast}$. For $J^{\ast} = 0.01$, for instance, 
Fig.~\ref{fig2}~(a) shows that the dependence of the density 
on the chemical potential is smooth for $T^{\ast}=0.14$, but for $T^{\ast}=0.11$ there is a 
discontinuity at $\mu^{\ast}=-1.679$. The same behavior persists until $J^{\ast} = 0.06$. 
A closer inspection of the configurations generated at this chemical 
potential, Fig.~\ref{fig3}, reveals an order-disorder phase transition. 
In this regime the electrostatic interaction drives the phase separation and only a 
tricritical point is observed. The effect of the short-range repulsion can be noticed 
only as a decrease of the tricritical temperature and increase of the 
corresponding density, as shown in Fig.~\ref{fig4}, our estimate for the 
phase diagram as a function of the repulsive strength $J^{\ast}$, obtained 
from histogram reweighting. We give an estimate of the location of the tricritical point, 
as shown in Fig.~\ref{fig4}, based on a linear extrapolation of the coexisting lines, as 
expected for $d=3$ tricriticality. We did not find any evidence of a gas-liquid 
phase separation for $J^{\ast}$ between 0 and 0.06. 

When we increase the repulsive strength $J^{\ast}$ the nearest neighbor 
occupancy becomes less favorable. Consider the reduced electrostatic energy 
defined as follows:
\begin{equation}
U_{ij}^{\ast}=\frac{U_{ij}}{E_0}=\frac{z_i z_j}{r_{ij}^{\ast}}\;,
\end{equation}
where $r_{ij}^{\ast}=r_{ij}/\sigma$ is the reduced separation between 
two ions of valences $z_i$ and $z_j$. In the lattice model depicted in 
Fig.~\ref{fig1} the distance between the nearest neighbor sites is $l$, 
while for the second nearest neighbor ones is $\sqrt{2}l$, where $l$ is the 
characteristic length of the simple cubic grid we are using in the simulations. 
Since the lattice RPM model corresponds to $l=\sigma$, if we consider $\sigma =1$, 
the energy between two ions in the first nearest neighbor sites is 
$z_i z_j + J^{\ast}$, while for the second nearest neighbor ones is 
$z_i z_j/\sqrt{2}$. Therefore, for $J^{\ast}=1-1/\sqrt{2}\approx 0.3$, if the site 0 in 
Fig.~\ref{fig1} is populated by a negative ion ($z_i = -1$), a 
positive ion ($z_j = +1$) will be affected by the same energy $-1/\sqrt{2}$ and 
can occupy first and second nn sites with equal probabilities. 
Therefore, we expect competition between first and second nearest neighbor 
occupancy in the region around $J^{\ast}=0.3$ . 
In the limit of $J^{\ast}\to \infty$ the occupation of the first nn sites is prohibited. 

\begin{figure}
\includegraphics[width=3.5cm]{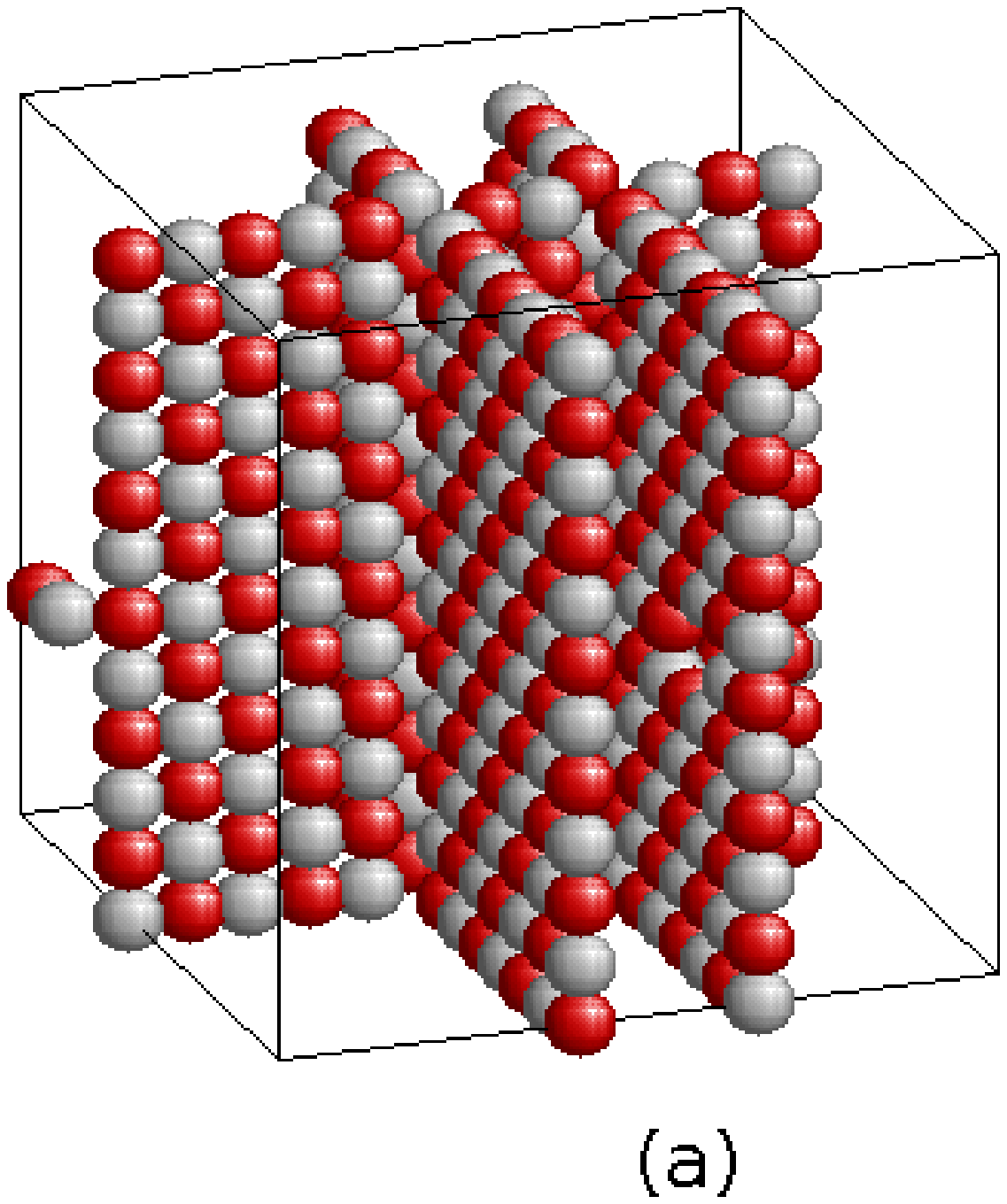}{\hspace*{.8cm}}
\includegraphics[width=3.5cm]{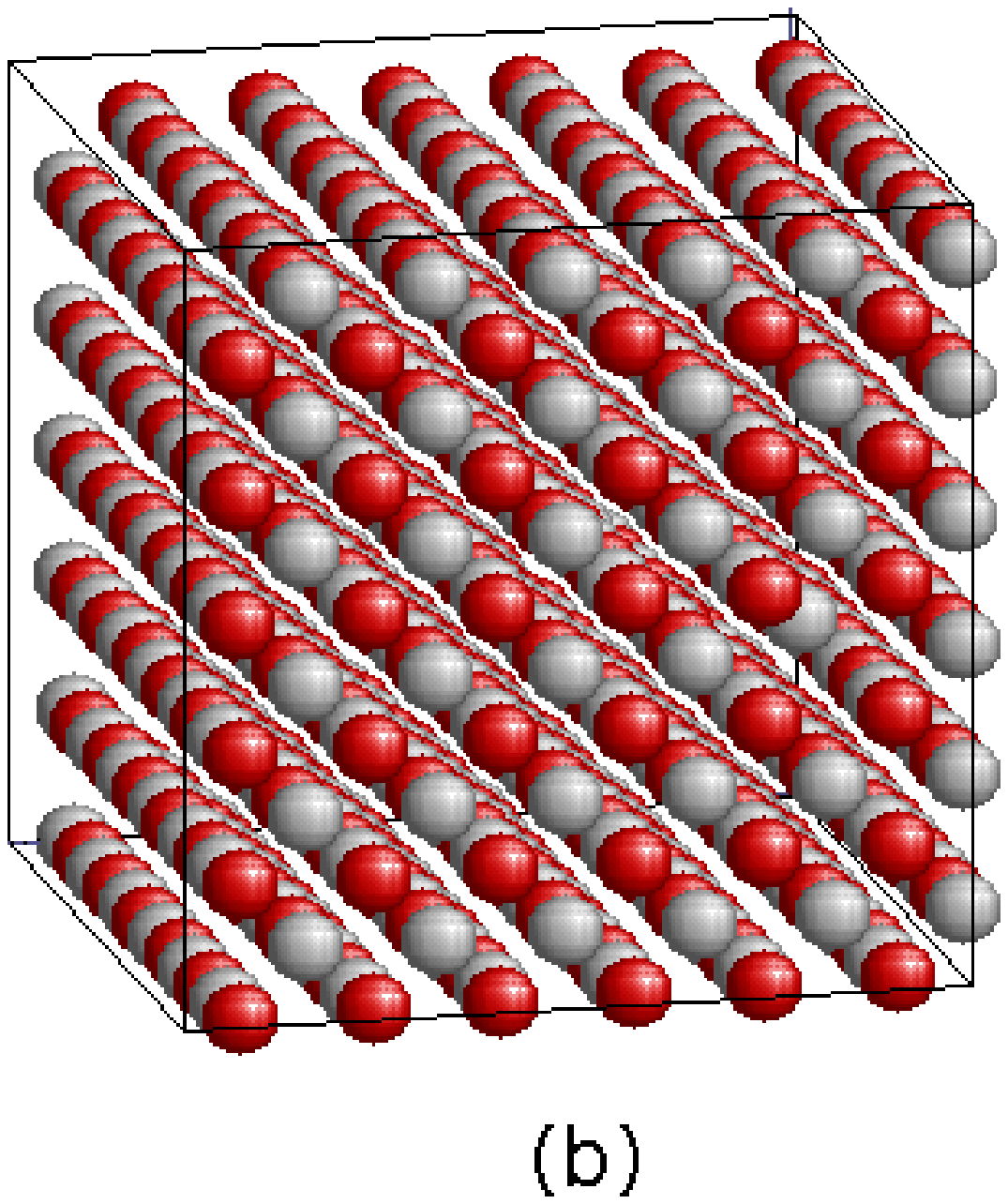}
\caption{\label{fig5} (color online) Configurations generated in the 
intermediate density region for (a) $J^{\ast}=0.07$ and (b) 0.1. 
In (a) the density is $\rho^{\ast}=0.17$ and 
the temperature is $T^{\ast}=0.04016$, while in (b) $\rho^{\ast}=0.5$ 
and $T^{\ast}=0.06$.} 
\end{figure}

\begin{figure}
\includegraphics[width=7cm]{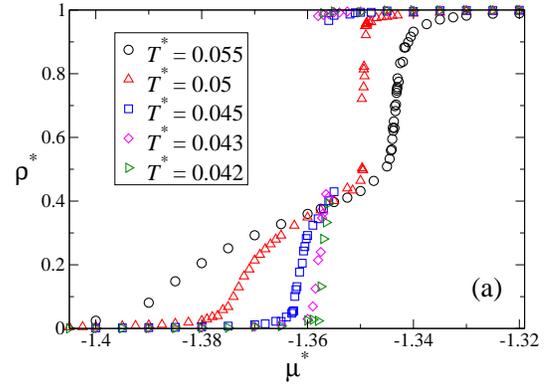}{\vspace*{.8cm}}
\includegraphics[width=7cm]{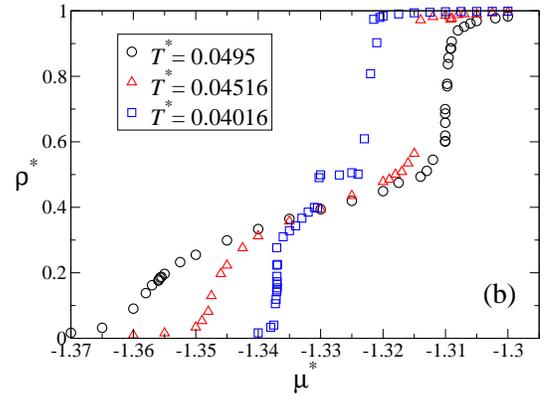}{\vspace*{.8cm}}
\includegraphics[width=7cm]{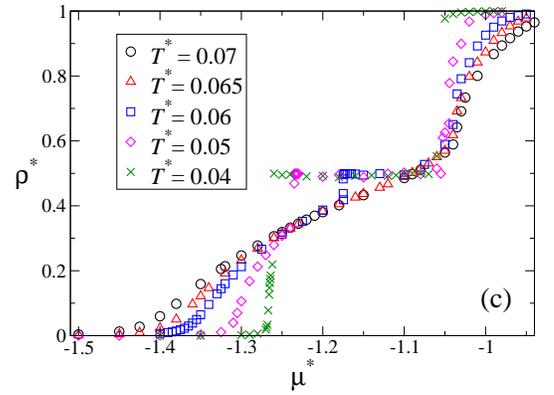}
\caption{\label{fig6} (color online) Isotherms calculated in grand canonical 
simulations ($L^{\ast}=12$) for different repulsive strength, 
(a) $J^{\ast}=0.065$, (b) 0.07 and (c) 0.1. 
Statistical uncertainties are smaller than the symbol size.} 
\end{figure}
In fact, we have found that above $J^{\ast}=0.065$ the phase diagram starts to change. 
While in Fig.~\ref{fig2} we have only a first-order phase transition between a low 
density disordered state and a high density antiferromagnetically ordered state, 
the simulations suggest the existence of structured configurations in the 
intermediate density region, between the disordered and antiferromagnetically ordered states. 
In Fig.~\ref{fig5} we show some of these configurations. We can see in 
Fig.~\ref{fig6} that there is a range of chemical potentials over which 
a plateau at $\rho^{\ast}=0.5$ starts to evolve, where the configurations are 
similar to that of Fig.~\ref{fig5}~(b). 

A run initiated either at low density disordered state or at a high density 
ordered state [see e.g. Figs.~\ref{fig3}~(a) and (b)] 
it would convert to this plateau. Although hysteresis loops are expected whenever 
first-order phase transitions are present, our GCMC method is unable to characterize these 
configurations properly. Configurations similar to Fig.~\ref{fig5}~(b) remain apparently 
stable even after 1.8$\times 10^9$ Monte Carlo steps, with a very low acceptance for the 
Monte Carlo moves ($< 0.01 \%$), due to the low temperature and high density. 
Such structures could represent metastable states, but we cannot determine this 
conclusively based on our simulations.     

\begin{figure}
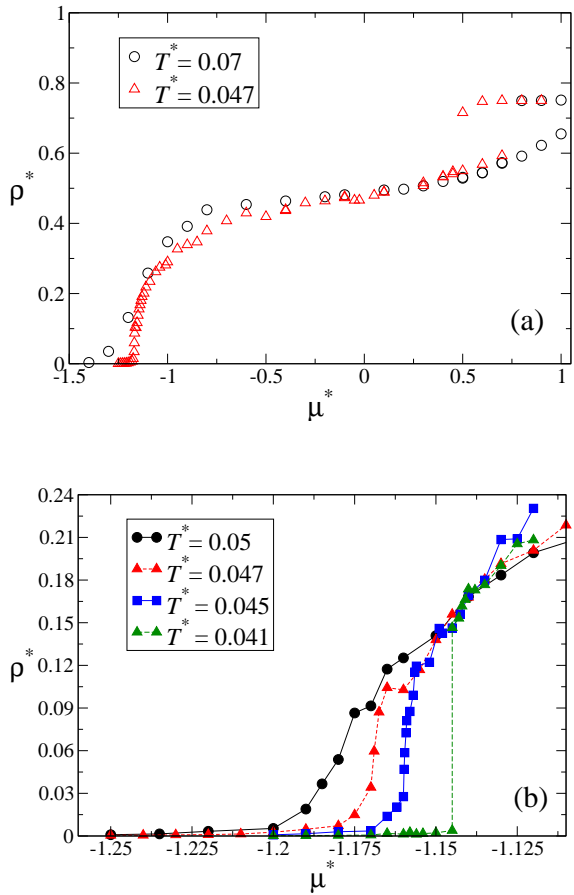

\includegraphics[width=7.5cm]{fig7a.eps}{\vspace*{.8cm}}
\includegraphics[width=7.5cm]{fig7b.eps}
\caption{\label{fig7} (color online) Density versus chemical 
potential for $J^{\ast}=0.3$. Statistical 
uncertainties are smaller than the symbol size.}
\end{figure}

For $J^{\ast}=0.3$, on the other hand, the simulations have produced a much 
clearer picture, as shown in Fig.~\ref{fig7}. Once again, hysteresis 
loops were observed but now for both low and high-density regions. 
Also, the plateau observed at $\rho^{\ast}=0.5$ disappears completely. 
For the low-density region Fig.~\ref{fig7}~(b) shows that the phase 
separation is between two disordered phases which we identify as a gas-liquid 
phase transition. 

In order to characterize the critical point associated with the gas-liquid 
transition, we have used FSS~\cite{Wilding92} analysis, 
which accounts for the lack of symmetry between coexisting phases in fluids. 
Briefly stated, for one-component systems we define an ordering parameter 
$M = N -sU$, where $s$ is the field-mixed parameter, 
such that at criticality the normalized probability distribution at a 
given system size, $P_{L}(x)$, has a universal form for every fluid in a given
universality class, with $x=A(M-M_c )$. As stated in Section ~\ref{model}, 
we did not attempt to incorporate pressure mixing in the scaling fields. 

\begin{figure}
\includegraphics[width=8cm]{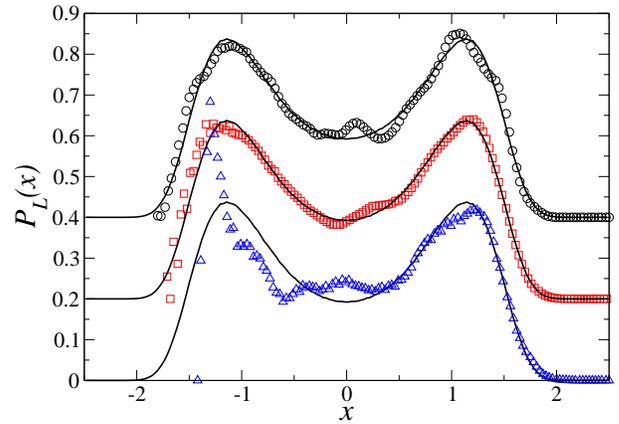}
\caption{\label{fig8} (color online) Ordering operator distribution 
for $\zeta = 1$ and $J^{\ast}=0.3$, for $L^{\ast} = 12$ (triangles), 
$L^{\ast} = 15$ (squares) and $L^{\ast} = 18$ (circles). 
The $L^{\ast}=15$ and $18$ curves have been displaced vertically for visual 
clarity. Lines are for the three-dimensional Ising 
universality class (data courtesy of N. B. Wilding).} 
\end{figure}

\begin{figure}
\includegraphics[width=8cm]{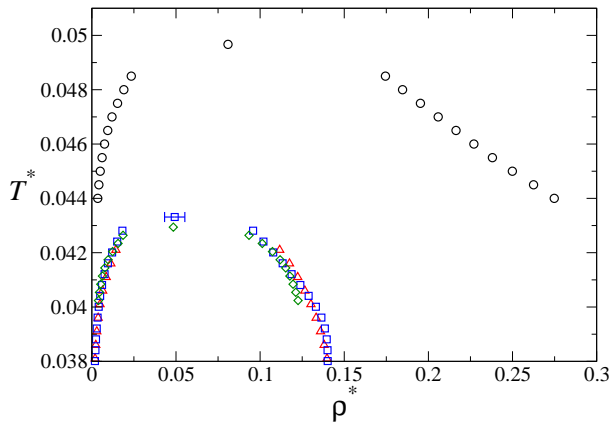}
\caption{\label{fig9} (color online) Gas-liquid coexistence curves 
for the lattice RPM model supplemented by a nearest neighbor repulsive strength 
$J^{\star}=0.3$, for $L^{\ast}=12$ (squares), 
$L^{\ast}=15$ (triangles) and $L^{\ast}=18$ (diamonds). The continuum RPM 
curves (circles) were taken from the finely discretized simulations from 
Panagiotopoulos and Kumar~\cite{azp99}. Statistical uncertainties smaller 
than the symbol size have been omitted.}
\end{figure}

In Fig.~\ref{fig8} we show the collapse 
of the measured $P_{L}(x)$ on the universal Ising ordering operator distribution 
for the system sizes $L^{\ast}=12$, 15 and 18. The quality of the collapse on 
the universal three-dimensional Ising critical distribution is better for the 
intermediate system size. We attribute the discrepancies for the larger system 
size to inadequate sampling, especially around the less probably states of $x=0$. 
For the smaller system size, there are too few particles in the simulation for 
adequate mapping on the universal distribution, as also observed previously for 
the RPM~\cite{azp02}. We suggest that the system presents critical behavior 
compatible with Ising-like behavior, with a critical point located at 
$T^{\ast}_c =0.04331\pm 0.00005$, $\rho^{\ast}_c = 0.049\pm 0.006$ and 
$\mu^{\ast}_c = -1.1537\pm 0.0005$ for a system size $L^{\ast}=15$. 
The critical temperature is reduced relative to the continuum RPM 
estimate~\cite{azp02}, $T^{\ast}_c = 0.0489\pm 0.0003$, mainly due to 
the addition of nearest neighbor repulsion in our model. The critical density, 
on the other hand, is considerable lower than $\rho^{\ast}_c = 0.076\pm 0.003$, 
the continuum RPM estimate~\cite{azp02}. 
In Fig.~\ref{fig9} we show the gas-liquid coexistence curves for the 
system sizes used in Fig.~\ref{fig8}, along with the continuum RPM 
result obtained from the finely discretized simulations of 
Panagiotopoulos and Kumar~\cite{azp99}.  
   
\begin{figure}
\includegraphics[width=4cm]{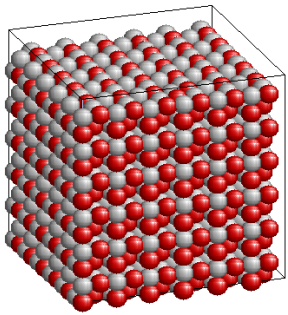}
\caption{\label{fig10} (color online) Typical configuration observed in 
the high-density region for $J^{\ast}=0.3$. The temperature is 
$T^{\ast}=0.07$ and the density is $\rho^{\ast}=0.75$.} 
\end{figure}

For the high-density region Fig.~\ref{fig7}~(a) suggests a first-order phase 
transition between a disordered high-density phase and an ordered state. 
Since these hysteresis loops still persist for a high temperature, 
we propose that the order-disorder phase transition with a tricritical point, 
observed for the low $J^{\ast}$ region, is replaced by a usual continuum-space 
behavior (first-order phase transition) as one increases $J^{\ast}$. In 
Fig.~\ref{fig10} we show a typical configuration for this high density region. Since 
we are increasing the first nn repulsion, the system evolves to a less 
denser configuration, instead of the antiferromagnetically ordered state observed 
for the low $J^{\ast}$ domain. This behavior is essentially the same of the 
theoretical predictions of Ciach and Stell's approach (see Fig.~1~(d) of Ref.~\cite{Ciach03}). 
We did not include the high density region for $J^{\ast}=0.3$ in Fig.~\ref{fig4} 
since GCMC simulations are unable to produce adequately sampled states at this very 
high densities.

From the phase diagram presented in Fig.~\ref{fig4} we can predict how the 
line of continuous order-disorder transitions present for the low nn repulsion 
changes into the continuum-space behavior as one increases $J^{\ast}$. For 
the low $J^{\ast}$ region, the gas-liquid transition remains metastable into the 
two-phase region associated with the order-disorder phase separation. When the 
nn strength is increased, the gas-liquid critical temperature increases, while the 
stable tricritical point decreases. Eventually, these two temperatures become of 
the same order, and the metastable critical point approaches the coexistence line 
of the order-disorder phase separation. This is essentially the behavior observed 
in Fig.~\ref{fig6}. From this point, the increase of $J^{\ast}$ makes the 
tricritical point metastable. 

Recently Ciach and Stell~\cite{Ciach03} have 
proposed a model where the same nn repulsion was added to the lattice RPM model. 
Using a model based on a Landau-Ginzburg-Wilson field-theoretical approach they 
have found a phase diagram very similar to Fig.~\ref{fig4} in the limit 
of strong nn repulsion. Also, they have obtained the evolution of the phase 
diagram from order-disorder to a fluctuation-induced first-order 
charge-ordered-charge-disordered transition for high densities~\cite{Ciach04}. 
Although GCMC is unable to properly characterize high-density phases, our 
simulation results predict that the sequence of phase diagrams observed in 
the Ciach and Stell's model is (a)$\to$(b)$\to$(d) in Fig.~1 of Ref.~\cite{Ciach03}, 
when we increase the nn strength. 

\section{Conclusions}
\label{conclusions}

In summary, we have used grand canonical Monte Carlo simulation
and histogram reweighting techniques to study phase transitions in
a lattice RPM model where, in addition to the Coulomb and
hard-core interactions, some short-range repulsive interactions
between the ions are added to the model. Phase diagrams for
different short-range strength have been obtained. Our simulation
results reveal a phase diagram strongly dependent on the nn
parameter $J^{\ast}$. Specifically, for weak nn repulsion and $\zeta =1$ only
order-disorder phase coexistence and a tricritical point are
observed, since the phase separation is driven by the
electrostatic interactions. Increasing $J^{\ast}$ makes the nearest neighbor 
occupancy becomes less favorable and the phase diagram starts to change. The 
nn exclusion on the simple cubic lattice used in our simulations prevents 
the order-disorder phase coexistence and tricriticality and for the 
low density region a gas-liquid phase separation appears. The critical 
point was estimated to belongs to the Ising universality class, with 
critical parameters at the same order of the continuum-space estimates. 
Thus our simulation results confirm qualitatively most of
the theoretical predictions of Ciach and Stell's approach~\cite{Ciach03,Ciach04}. 

\section*{Acknowledgments}
AD acknowledges financial support of the Brazilian agency 
CNPq - Conselho Nacional de Desenvolvimento Cient\'\i fico e Tecnol\'ogico. 
Funding by the Department of Energy, Office of Basic Energy Sciences (through
Grant No. DE-FG02-01ER15121 to AZP) and ACS-PRF (Grant 38165 - AC9
to AZP) are also gratefully acknowledged.

\newpage


\end{document}